\documentclass{article}
\usepackage{spconf,amsmath,graphicx}
\usepackage{xcolor}

\usepackage{enumitem}
\setlist{nosep, leftmargin=14pt}

\usepackage{mwe} 

\usepackage{amsmath,amsfonts,bm}

\def\1{\bm{1}}

\def\va{{\bm{a}}}

\def\vk{{\bm{k}}}

\def\vq{{\bm{q}}}
\def\vr{{\bm{r}}}

\def\vu{{\bm{u}}}

\def\vx{{\bm{x}}}
\def\vy{{\bm{y}}}
\def\vz{{\bm{z}}}

\def\mA{{\bm{A}}}

\def\mC{{\bm{C}}}
\def\mD{{\bm{D}}}

\def\mF{{\bm{F}}}

\def\mP{{\bm{P}}}
\def\mQ{{\bm{Q}}}

\DeclareMathAlphabet{\mathsfit}{\encodingdefault}{\sfdefault}{m}{sl}
\SetMathAlphabet{\mathsfit}{bold}{\encodingdefault}{\sfdefault}{bx}{n}

\def\gE{{\mathcal{E}}}
\def\gF{{\mathcal{F}}}

\def\gL{{\mathcal{L}}}

\def\gP{{\mathcal{P}}}

\newcommand{\R}{\mathbb{R}}

\DeclareMathOperator*{\argmin}{arg\,min}

\newcommand{\T}{\top}

\title{Accelerated optimization of implicit neural representations for CT reconstruction}
\name{Mahrokh Najaf, Gregory Ongie}
\address{Marquette University, Department of Mathematical and Statistical Sciences, Milwaukee, WI, USA.}
\begin{document}
%
\maketitle
\begin{abstract}
Inspired by their success in solving challenging inverse problems in computer vision, implicit neural representations (INRs) have been recently proposed for reconstruction in low-dose/sparse-view X-ray computed tomography (CT). An INR represents a CT image as a small-scale neural network that takes spatial coordinates as inputs and outputs attenuation values. Fitting an INR to sinogram data is similar to classical model-based iterative reconstruction methods. However, training INRs with losses and gradient-based algorithms can be prohibitively slow, taking many thousands of iterations to converge. This paper investigates strategies to accelerate the optimization of INRs for CT reconstruction. In particular, we propose two approaches: (1) using a modified loss function with improved conditioning, and (2) an algorithm based on the alternating direction method of multipliers. We illustrate that both of these approaches significantly accelerate INR-based reconstruction of a synthetic breast CT phantom in a sparse-view setting.

\end{abstract}
\section{Introduction}
Machine learning (ML) has shown promise in solving many challenging ill-posed image reconstruction problems in X-ray computed tomography (CT). Much of the attention has been given to supervised or semi-supervised approaches, which require access to ``ground truth'' target images. However, in many imaging applications, there is limited access to fully-sampled and/or noise-free ground truth images. In these cases, unsupervised ML techniques, that operate without ground truth, may still be used.

Inspired by their success in solving challenging inverse problems in computer vision in an unsupervised manner, researchers have begun to use 
implicit neural representations (INRs) to solve inverse problems in CT reconstruction \cite{sun2021coil, shen2022nerp, song2023piner, saragadam2023wire, fu2024attenuation}. An INR encodes an image as a small-scale neural network (typically a multi-layer perceptron) that takes as input spatial coordinates, and outputs the corresponding grayscale value at that location. The advantage of INRs is that they can be far more parameter-efficient compared to traditional pixel/voxel representations. Additionally, INRs are a fully differentiable representation, and can be optimized with standard neural network training algorithms and software. Finally, INRs impart an implicit regularization effect that helps prevent overfitting to noise and artifacts.

Conceptually, INRs belong to the class of model-based iterative reconstruction (MBIR) methods. 

However, relative to traditional MBIR, fitting INRs in the context of CT imaging can be prohibitively slow, requiring many thousands of iterations to reach acceptable image quality.

To address this issue, we propose two approaches in this paper to accelerate the optimization of INRs in the context of CT reconstruction. First, we propose minimizing a modified least squares loss function that can be viewed as a pre-conditioning strategy. Second, we propose an INR fitting algorithm based on the alternating direction method of multipliers (ADMM). 

We illustrate both approaches in reconstructing simulated sparse-view breast CT data. Our experiments show substantial improvements in the speed of convergence and overall reconstruction quality across three different INR architectures relative to training with a standard least squares loss.

\section{Problem Formulation}
Suppose we are given sinogram measurements $\vy \in \mathbb{R}^m$ (re-arranged into a vector). Let $f_\theta : \R^d\rightarrow \R$ denote a given INR architecture with trainable parameters $\theta \in \R^p$. The goal is to find a set of INR parameters $\theta \in \R^p$ such that $\gP\{f_\theta\} \approx \vy$,
where $\gP$ is a linear operator approximating line integrals of the function $f_\theta$ over a set of lines dictated by the acquisition geometry. 

The standard approach to INR-based CT reconstruction is to pose this as an optimization problem:
\begin{equation}\label{eq:inrfit}
\min_{\theta} \gL(\gP\{f_\theta\},\vy),
\end{equation}
where $\gL$ is a differentiable loss function. A common choice is the least squares loss:
\begin{equation}\label{eq:LS}
\gL_{\text{LS}}(\vz,\vy) = \|\vz - \vy\|^2.
\end{equation}

When using INRs, there is flexibility in implementing $\gP$. For this work, we focus on a simple approach where the INR is rasterized prior to approximating line integrals. In symbols:
\[
\gP\{f_\theta\} = \mP \gE\{f_\theta\},
\]
where $\gE\{\cdot\}$ denotes the evaluation operator that converts an INR into a rasterized image, i.e., if $\vx = \gE\{f_\theta\}$, then $\vx = (f_\theta(\vr_{i}))_{i=1}^n \in \R^n$, where $\{\vr_{i}\}_{i=1}^n \subset \R^d$ are coordinates belonging to a regular spatial grid, and $\mP \in \R^{m\times n}$ is a matrix approximating line integrals of rasterized images.

Under this assumption, when using the least squares loss, the INR fitting problem \eqref{eq:inrfit} is equivalent to the (non-linear) least squares problem:
\begin{equation}\label{eq:inr_fit_ls}
\min_\theta \|\mP\mathcal{E}\{f_\theta\} - \vy\|^2.
\end{equation}
A fundamental challenge in efficiently optimizing \eqref{eq:inr_fit_ls} with gradient methods is that
the system matrix $\mP$ arising in CT applications is ill-conditioned. Below, we propose two alternatives to optimizing \eqref{eq:inr_fit_ls} that circumvent this ill-conditioning.

\section{Proposed Approaches}
\subsection{Filtered Least Squares Loss Function}

 To address the ill-conditioning arising when using a least squares loss, we propose using a modified least squares loss that incorporates an ``FBP-type'' preconditioning matrix. We note that such preconditiong strategies have a long history in MBIR for CT reconstruction; see \cite{wang2019fast} and references therein. 

In particular, let $\mF$ be a matrix representation of the filtering operation in filtered back-projection (FBP).
We consider the following modified loss, which we call filtered least squares (FLS):
\begin{equation}\label{eq:FLS}
\gL_{\text{FLS}}(\vz,\vy) = (\vz-\vy)^\top \mF (\vz-\vy).
\end{equation}
Note that if $\mF$ represents convolution with a ramp filter, as in classical FBP, then $\mF$ is a positive semi-definite matrix, since we can write $\mF$ as the Kronecker product of convolution matrices $\mC = {\bm\gF}^*\mD{\bm\gF}$, where $\bm\gF$ is the discrete Fourier transform, and $\mD$ is a diagonal matrix with non-negative values. Therefore, we can equivalently express the filtered least squares loss as
\[
\gL_{\text{FLS}}(\vz,\vy) = \|\mF^{1/2}(\vz-\vy)\|^2,
\]
where $\mF^{1/2}$ is a positive semi-definite matrix satisfying $\mF = (\mF^{1/2})^\T \mF^{1/2}$. 
Again, in the case that $\gP\{f_\theta\} = \mP\mathcal{E}\{f_\theta\}$, and using the filtered least squares loss in \eqref{eq:inrfit}, we arrive at the non-linear least squares problem:
\[
\min_\theta \|\mF^{1/2}(\mP\mathcal{E}\{f_\theta\} - \vy)\|^2.
\]

\begin{table}[]
    \centering
    \begin{tabular}{r|c} & condition number ratio: 
$\kappa_{\text{FLS}}/\kappa_{\text{LS}}$\\
            \hline
        ReLU & $(3.80 \pm 1.52) \times 10^{-3}$\\
        SIREN & $(1.51 \pm 0.01) \times 10^{-1}$ \\
        Hash Enc. & $(2.21\pm 1.41) \times 10^{-3}$
    \end{tabular}
    \caption{\small Ratio of condition numbers of the Gram matrix associated with least squares ($\kappa_{\text{LS}}$) and filtered least squares ($\kappa_{\text{FLS}}$) at initialization, using three different INR architectures (ReLU, SIREN, Hash Enc.). We report the mean $\pm$ std. dev. over 100 random INR parameter initializations.}
    \label{tab:cond_num}
\end{table}
\subsubsection{Linearized analysis}
Suppose our INR architecture is expressible as $f_\theta(\vr) = \sum_{i=1}^W a_i f_i(\vr;\theta')$ where $\va = [a_1,...,a_W]$ are the final-layer network weights, and $\theta'$ is the collection of all parameters excluding the outer-layer weights. Then if $\vx = \mathcal{E}\{f_\theta\}$, by linearity of the evaluation operator $\mathcal{E}$, we have
\[
\vx = \sum_{i=1}^W a_i \mathcal{E}\{f_i(\cdot,\theta')\} = \mQ\va,
\]
where $\mQ = \mQ(\theta')$ is the $n\times W$ matrix whose $i$th column is $\mathcal{E}\{f_i(\cdot,\theta')\}\in \R^n$; we call this the INR feature matrix. Assuming the feature matrix $\mQ$ is fixed, and only the final-layer weights $\va$ are optimized, substituting this into \eqref{eq:inr_fit_ls} gives:
\[
\min_{\va\in\R^W} \|\mP\mQ\va - \vy\|^2.
\]

This is a linear least squares problem with respect to the matrix $\mA = \mP \mQ \in \R^{m\times W}$. 
It is well-known that the convergence rate of gradient methods for solving least squares depends on the condition number $\kappa$ of the Gram matrix  $\mA^\T\mA$ \cite{beck2014introduction}. Some strategies have been proposed to improve the conditioning of the INR feature matrix $\mQ$. For example, changing the activation function from the ReLU to a B-spline wavelet activation function was shown to improve the conditioning of $\mQ$ in \cite{shenouda2024relus}.
Nevertheless, due to the presence of $\mP$, the matrix $\mA^\T\mA = \mQ^\T\mP^\T\mP\mQ$ is still ill-conditioned regardless of any architectural changes that improve the conditioning of $\mQ$.

Repeating the analysis above for the filtered least squares loss \eqref{eq:FLS} in place of the least squares loss \eqref{eq:LS}, we see that the convergence rate of the corresponding linear least squares approximation depends on the condition number of the matrix $\mQ^\T\mP^\T\mF\mP\mQ$. In particular, the conditioning of the inner matrix $\mP^\T\mF\mP$ is improved, since it acts as an approximate FBP. 

In Table \ref{tab:cond_num}, we empirically verify this claim by comparing the condition number of the Gram matrix arising from the least squares loss versus the filtered least squares loss. In particular, we consider the INR feature matrices $\mQ$ arising from three different INR architectures \emph{at initialization} (see Section \ref{sec:exp} for more details on these architectures), and compare the ratio of the condition numbers. We find that the condition number corresponding to the filtered least squares loss is 1--3 orders of magnitude smaller (on average), which suggests the filtered least squares loss should be substantially faster to optimize with gradient methods in the early stages of optimization across all of these INR architectures.

\subsection{ADMM Algorithm}
 Instead of directly altering the loss function $\gL$, which may be undesirable from a statistical modeling point of view (e.g., interpreting the loss as the negative log-likelihood), we propose fitting an INR with a standard loss, but using an alternative optimization algorithm. 
 In particular, we focus on optimizing the INR fitting problem  with least squares loss \eqref{eq:inr_fit_ls}.

 An equivalent constrained form of \eqref{eq:inr_fit_ls} is given by
\begin{equation}\label{eq:eq_const_ver}
\min_\theta \tfrac{1}{2}\|\mP\vx -\vy\|^2~~s.t.~~\vx = \gE\{f_\theta\}.
\end{equation}
To optimize \eqref{eq:eq_const_ver}, we propose applying the alternating direction method of multipliers (ADMM) algorithm \cite{boyd2011distributed}. The (rescaled) augmented Lagrangian associated with \eqref{eq:eq_const_ver} is 
\[
AL(\vx,\theta, \vu, \mu) = \tfrac{1}{2}\|\mP\vx -\vy\|_2^2 + \tfrac{\mu}{2}\|\vx - \gE\{f_\theta\} + \vu\|_2^2,
\]
where $\vu \in \R^n$ has the interpretation as a vector of Lagrange multipliers. This gives the ADMM updates: starting from initializations $\vq_0$, $\vu_0$, for $k=0,1,2, ...$, set
\begin{align}
\vx_{k+1} & = \argmin_{\vx} \tfrac{1}{2}\|\mP\vx -\vy\|_2^2 + \tfrac{\mu}{2}\|\vx - (\vq_k - \vu_k)\|_2^2 \label{eq:xup}\\
\theta_{k+1} &  = \argmin_\theta \|\gE\{f_\theta\} - (\vx_{k+1} + \vu_k)\|_2^2 \label{eq:thetaup}\\
\vq_{k+1} & = \gE\{f_{\theta_{k+1}}\}\\
\vu_{k+1} & = \vu_k + \vx_{k+1} - \vq_{k+1}.
\end{align}
The $\vx$-update \eqref{eq:xup} is a linear least squares problem, and can be solved with an efficient iterative method, such as the conjugate gradient least squares (CGLS) algorithm \cite{paige1982lsqr}. The $\theta$-update \eqref{eq:thetaup} is an INR training problem. However, in \eqref{eq:thetaup}, the INR fitting takes place in pixel space, and does not require costly forward/backward projections in each iteration, unlike directly optimizing the least squares loss \eqref{eq:LS}. For computational efficiency, we propose to use inexact solves of both these subproblems.

 Since the constraint $\vx = \mathcal{E}(f_\theta)$ is nonlinear, the iterates of ADMM are not guaranteed to converge. Nonetheless, convergence can be monitored by measuring the primal and dual residuals as proposed in \cite{wang2017nonconvex}.

\begin{figure*}[!htb]
    \centering
     \vspace{-4.5em}
    \includegraphics[width=\textwidth]{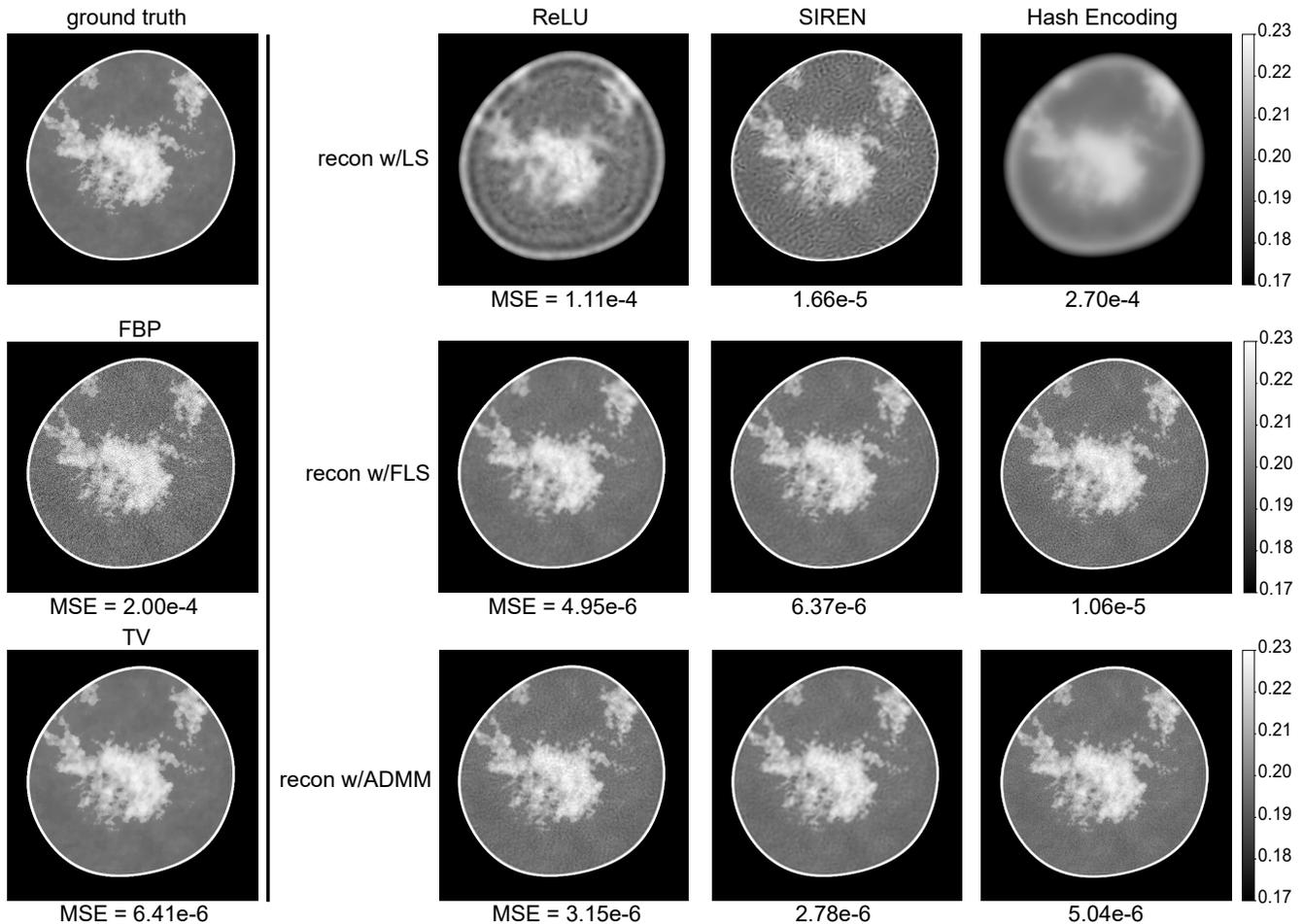}
    \caption{Reconstruction of simulated sparse-view breast CT data with various INR architectures (ReLU, SIREN, Hash Encoding) using a standard least squares (LS) loss, the proposed filtered least squares (FLS) loss, and the proposed ADMM algorithm. The left panel shows a baseline filtered back-projection (FBP) and total variation (TV) regularized reconstruction. } 
    \label{fig:main}
\end{figure*}

\section{Experiments}\label{sec:exp}
We illustrate the proposed INR training modifications for the reconstruction of a breast CT phantom from simulated noisy sparse-view data. We use a modified version of the phantom introduced in \cite{phantom}; the phantom is modified to have texture, and a smooth deformation is applied such that the boundary of the breast is non-circular. Sinogram data of the phantom is simulated under a 2D circular, fan-beam scanning geometry, which is representative of the mid-plane slice of
a 3D circular cone-beam scan. 

We simulate 128 equally distributed views over a 360 degree scan and the projections are sampled on a 1024-pixel detector. Poisson noise is simulated in transmission space assuming $32\times 10^{10}$ total incident photons, or approximately $2.4\times 10^6$ incident photon per detector pixel. To avoid the inverse crime, we simulate projections from the phantom rasterized on a $2048\times2048$ pixel grid. We take as ground truth the phantom downsampled to a $512\times 512$ pixel grid. For all reconstructions, we use discrete 2D fan-beam projection operators operating on a $512\times 512$ pixel grid.

As baselines, we compare with an FBP reconstruction using a ramp filter, and a total variation (TV) regularized least squares reconstruction which is optimized using the Chambolle-Pock primal-dual algorithm \cite{sidky2012convex}. We tune the regularization strength to achieve the smallest pixel-wise mean-squared-error (MSE), defined as $\text{MSE} = \tfrac{1}{n}\|\vx-\vx^*\|^2,$ where $\vx$ is the reconstructed image and $\vx^*$ is the ground truth image, and $n = 512^2$ is the number of pixels.

For the INR reconstructions, we investigate three different architectures: ReLU networks with an initial Fourier features embedding \cite{tancik2020fourier}, SIREN networks \cite{sitzmann2020implicit}, and Hash Encoding networks \cite{muller2022instant}. We use the following settings:
\begin{itemize}[leftmargin=*]
\item \textbf{ReLU:} depth 6, 256 hidden features in each layer, and Fourier features with frequencies $2\pi\vk$, $\vk \in \mathbb{Z}^2$, $\|\vk\|\leq 15$ (assuming INR is defined over coordinates in $[0,1]^2$).
\item \textbf{SIREN}: depth 6, 256 hidden features in each layer, with base frequency $\omega_0 = 75$ in all layers.
\item \textbf{Hash Encoding}: size of the hash table $T=23$, levels $L=16$, number of feature dimensions per entry $F=2$, min/max level resolutions $N_{\min} = 16$ and $N_{\max} =256$, and depth 6 MLP with 128 hidden features per layer.
\end{itemize}

When fitting INRs with the LS or FLS loss functions, we use the Adam optimizer. All networks are trained with the same learning rate schedule: 500 iterations with a learning rate of $\tau_0$ followed by 500 iterations with a learning rate of $\tau_0/10$. For ReLU INR, we use $\tau_0 = 1\times 10^{-3}$, and for Hash Encoding and SIREN INRs, we use $\tau_0 = 1\times 10^{-4}$. 

When fitting INRs with the ADMM algorithm, we use 20 outer ADMM iterations. To approximate a solution to the INR fitting subproblem, we use 50 iterations of Adam with a fixed learning rate of $1\times 10^{-4}$ for ReLU/SIREN and $1\times 10^{-3}$ for Hash Encoding. We also use 50 iterations of CGLS to approximate a solution to the least squares problem. The ADMM parameter $\mu$ is tuned as follows: ReLU $\mu = 1$, SIREN $\mu = 3$, Hash Encoding $\mu = 2.5$.

Figure \ref{fig:main} shows reconstructions obtained after 1000 iterations\footnote{Here ``iteration'' is understood as one matrix-vector product each with $\mP$ and $\mP^\T$; this is equivalent to one gradient step for the LS and FLS approaches, and one iteration of CGLS for ADMM.} of each approach. We observe that all three INR architectures yield poor quality reconstructions when trained with the LS loss. Reconstruction trained with the FLS loss and the ADMM approach give higher quality reconstructions that are visually similar to the TV baseline and have smaller or comparable MSE.

\begin{figure}[!htb]
\includegraphics[width=\columnwidth]{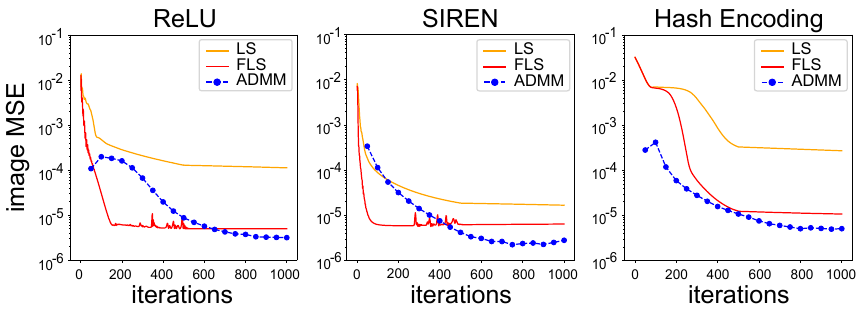}
\vspace{-1.5em}
    \caption{\small Reconstruction MSE vs iterations for the different INR architectures optimized using a standard least squares (LS) loss, the proposed filtered least squares (FLS) loss, and the proposed ADMM algorithm.}
    \label{fig:main2}
\end{figure}

In Figure \ref{fig:main2}, we compare the reconstruction MSE over 1000 iterations. Both FLS and ADMM substantially improve over LS in terms of the speed of reduction of image MSE. Training with FLS gives the fastest reduction in image MSE for ReLU and SIREN architectures, while ADMM gives a slightly faster reduction in MSE for the Hash Encoding architecture. However, the ADMM approach gives the lowest final MSE in all cases. 

\vspace{-0.2em}

\section{Discussion and Conclusion}
This work proposed using a modified loss function (filtered least squares) and a modified training algorithm (nonlinear constrained ADMM) for efficient training of INRs in CT reconstruction. Our results on simulated breast CT data show that both approaches substantially accelerate INR optimization relative to directly minimizing a least squares loss with a gradient-based optimizer.
Another common strategy to accelerate optimization in CT reconstruction is to use ordered subsets techniques \cite{hudson1994accelerated}. While not investigated in this work, incorporating ordered subsets into the proposed approaches could further accelerate the optimization of INRs. Recent work has also explored a meta-optimization technique for fast INR training \cite{tancik2021learned}. This and other semi-supervised approaches are complementary to the proposed approaches, and their synthesis is  an interesting direction for future work.

\newpage
\section{Compliance with ethical standards}
\label{sec:ethics}
This is a numerical simulation study for which no ethical approval was required.

\section{Acknowledgments}
\label{sec:acknowledgments}
This work was supported by NSF CRII award CCF-2153371.


\bibliographystyle{IEEEbib}
\bibliography{strings,refs}

\end{document}